\documentclass[11pt,twoside,psfig]{article}

\usepackage{asp2006}
\usepackage{epsf}
\usepackage{lscape}

\newcommand{\tal}{\it et al. \rm}

\begin{document}
\title{Formation and evolution of bulges}   
\author{E. Athanassoula$^{1}$ and I. Martinez-Valpuesta$^{1,2}$}   
\affil{$^1$LAM, OAMP, 2 place Le Verrier, 13248
  Marseille cedex 04, France\\
$^2$IAC, E-38200 La Laguna, Tenerife, Spain\\}    

\begin{abstract} 
After presenting three ways of defining a bulge component in disc
galaxies, we introduce the various types of bulges, namely the classical
bulges, the boxy/peanut bulges and the disc-like bulges. We then
discuss three specific topics linked to bulge formation and evolution,
namely the coupled time evolution of the bar, buckling and peanut
strengths; the effect of velocity anisotropy on peanut formation; and
bulge formation via bar destruction.
 \end{abstract}



\section{What is a bulge?}

Three ways of defining a bulge have been used so far, one morphological,
the second photometrical and the third kinematical. Based on
morphology, a bulge is a component of a disc galaxy that has a  
smooth light distribution that swells out of the central part of a
disc viewed edge-on. This definition has the disadvantage of being
applicable only to edge-on systems and the advantage of necessitating
only an image of the galaxy. The second definition is based on
photometry and defines a bulge as the extra light in the central part
of the galaxy, above the exponential profile fitting the remaining (non
central) part of the disc. In earlier papers this component was fitted
with an 
$r^{1/4}$ law, while more recent ones use its generalisation to an
$r^{1/n}$ law (S\'ersic 1968). This definition has the
advantage of being applicable to disc galaxies independent of their
inclination. It has also the advantage of leading
to quantitative results about the light distribution, but has the
disadvantage of assigning to the bulge any extra central luminosity of
the disc, independent of its origin. The third definition is based on 
kinematics, and in particular on the value of $V/\sigma$, or more
specifically on the location of the object on the ($V/\sigma$,
ellipticity) plot, which is often referred to
as the Binney diagram (Binney 1978, 2005). 
This definition,
potentially quite powerful, has unfortunately been very little used
so far, due to the small number of galaxies for which the necessary
data are available, a situation which is rapidly improving
with large surveys, such as SAURON (Bacon \tal 2001; 
de Zeeuw \tal 2002; Peletier this volume). 

The lack of a single, clear-cut definition of a bulge is due to the
fact that disc galaxies are viewed in different orientations and also
to the fact that not all types of data are available for all objects. 
Nevertheless, it has led to considerable confusion and to the fact
that bulges are an inhomogeneous class of objects. Indeed, many
different types of objects, with very different properties and
formation histories are included in the general term
`bulges'. To remedy this, Kormendy (1993) and Kormendy \& Kennicutt (2004)
distinguish classical bulges from pseudo-bulges, the latter category
encompassing all bulges that are not classical. Athanassoula (2005a)
argues that pseudo-bulges also are an inhomogeneous class of objects,
and thus distinguishes three types of bulges. 
 
{\bf Classical bulges} are formed by gravitational collapse or 
hierarchical merging of smaller objects and corresponding dissipative gas
processes. The material forming this bulge could be externally
accreted, or could come from clumps in the proto-disc. In general,
bulges of this type are formed before the actual disc (e.g. Steinmetz \&
M\"uller 1995; Noguchi 1998; Immeli \tal 2004). Nevertheless, a bulge can
also form from material externally accreted at much later stages
(e.g. Pfenniger 1991; Athanassoula 1999; Aguerri, Balcells \& Peletier
2001; Fu, Huang, 
Deng 2003). Their morphological, photometrical and kinematical properties
are similar to those of ellipticals.

{\bf Box/peanut bulges} (B/P) form from a vertical instability of 
the disc material. This has often been observed in $N$-body
simulations of bar-unstable discs, where the initial stage of
bar formation is followed by a puffing up of the inner parts of the
bar (e.g. Combes \& Sanders 1981; Combes \tal 1990; Raha \tal 1991;
Athanassoula \& Misiriotis 2002; Athanassoula 2003, 2005a; O'Neil \&
Dubinski 2003; Martinez-Valpuesta \& Shlosman 2004; Debattista \tal
2004, 2006; Martinez-Valpuesta \tal 2006).
Viewed side-on (i.e. edge-on with the line of sight along the bar
minor axis), this structure protrudes from the disc and has a 
characteristic boxy or peanut shape whose size is of the order of a
few disc scale-lengths. Thus, a box/peanut bulge is just {\it part} of a bar 
seen side-on. 

Finally {\bf disc-like bulges} form from inflow of
(mainly) gas material to the centre of the galaxy and subsequent star
formation (e.g. Athanassoula 1992; Friedli \& Benz 1993; Heller \&
Shlosman 1994; Wada \& Habe 1995) . The torque exerted by the bar
pushes gas, and to a lesser extent 
also stars, to the inner parts of the disc where they 
form an inner disc. Star formation can be very high there, due to
the increased gas density. Thus the result of this process should be a
central disc, or disc-like object, whose stellar component should
include a sizable fraction of young stars and whose size should be
less than, or of the order of a kpc. Such a component could harbour
sub-structures such as spirals, or bars, as is indeed sometimes
observed (Kormendy \& Kennicutt 2004 and references therein). It is
thus clear that disc-like 
bulges are very different objects from boxy/peanut bulges, since they
are much smaller, have a different shape, different kinematics and
provide a different type of excess on the radial photometric profiles.  
They also have different formation histories.

The different formation histories of these three types of objects lead
to different properties -- morphological, photometrical and kinematical --
which in turn help classify observed bulges into one of the three
above mentioned types. Nevertheless, as stressed by Athanassoula
(2005a), different types of bulges often co-exist and it is 
possible to find all three types of bulges in the same simulation,
or galaxy. 

Realising the non-homogeneity of the objects classified as bulges and
attempting to classify them is only the first step. Much more work is
now necessary, particularly on two issues. The first one is the
understanding of the formation and evolution of these types of
objects. The second one is to predict the properties of these objects,
starting from their formation scenarios. The latter is particularly
important in order to bridge the gap between classification schemes based on
formation histories and classification schemes based on observed
properties. Here we make small contributions to both these
issues, using $N$-body simulations. In Sect. 2 we present the
time evolution of the bar, the buckling and the peanut strengths and
their interplay. Sect. 3 discusses the velocity anisotropy and its
link to the above strengths. Finally, in Sect. 4 we discuss the
photometrical properties of a destroyed bar and boxy/peanut bulge.

\section{Time evolution of the bar, buckling and peanut strengths}

Formation of boxy/peanut bulges has been witnessed in a large number
of simulations, starting with Combes \& Sanders (1981). It is now well
understood in terms of orbital structure and particularly in terms of
the instability of the periodic orbits that constitute
the backbone of the bar (Pfenniger 1984; Skokos, Patsis \&
Athanassoula 2002; Patsis, Skokos \& Athanassoula 2002).  

\setcounter{figure}{0}
\begin{figure}[!h]
\plotfiddle{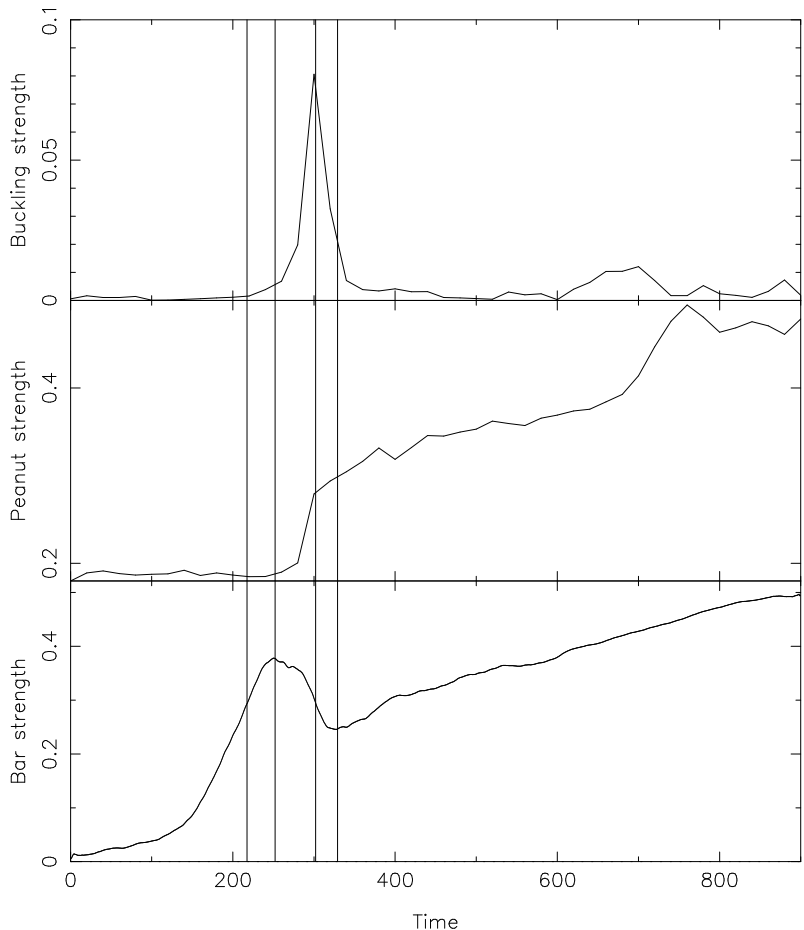}{12.cm}{0}{70}{70}{-200}{-150}
\caption{Time evolution of three peanut-, or bar-related quantities,
  namely the buckling strength (asymmetry; upper panel), the peanut
  strength (middle panel) and bar strength (lower panel). The thin
  vertical lines mark characteristic times linked to bar
  formation and evolution. From left to right, these are the bar
  formation time, the maximum amplitude time, the bar decay time and
  the bar minimum amplitude time (see text).} 
\label{fig:tevl} 
\end{figure}

Time evolution of the bar, the buckling and the peanut strengths 
are plotted in Fig.~\ref{fig:tevl} for a simulation which develops a
strong bar. The time is given
in computer units and, for a reasonable calibration, 100 computer units
correspond to 1.4 Gyrs (Athanassoula \& Misiriotis 2002).
The initially unbarred disc forms a bar roughly between times 150 and
250 (lower panel). We define as bar formation time the time at which
the bar-growth 
is maximum (i.e. when the slope of the bar strength as a function of
time curve is maximum) and indicate it by the first vertical line in
Fig.~\ref{fig:tevl}. The bar strength reaches a maximum at a time
noted by the second vertical line, and then drops, at a time which is
denoted by the third vertical line. It reaches a minimum, at a time
given by the fourth vertical line and then starts increasing
again. The upper panel shows the buckling strength, i.e. the vertical
asymmetry as a function of time. Before the bar forms, the disc is
vertically symmetric, with the first indications of asymmetry occurring
after bar formation. Then the asymmetry grows very abruptly to a
strong, clear peak and then drops equally abruptly. The time of the
buckling is given by the peak of this curve and is very clearly
defined. It is important to note that, to within the measuring errors,
it coincides with the 
time of bar decay (third vertical line). This is not accidental. We
verified this in a very large number of simulations and thus can
establish the link between the buckling episode and the decay of the bar
strength (Raha \tal 1991; Martinez-Valpuesta \& Shlosman
2004). The middle panel shows the strength of the peanut as a
function of time. This quantity grows abruptly with time after the bar
has reached its maximum amplitude and during the time of the buckling.
This abrupt growth is followed by a much slower increase over a long
period of time. Taken together, these figures show that the bar forms
vertically thin, and only after it has reached a maximum strength does the
buckling phase occur. The time intervals during which bar
formation, or peanut formation, or buckling occur are all three rather
short, of the order of a Gyr or less, and they are followed by much longer
times of slower bar and B/P evolution.     

This particular simulation has a second, much weaker buckling episode
around time 
700. This occurs very often in simulations developing strong
bars. It is seen clearly in the peanut strength development, as a second
abrupt increase of the peanut strength (Athanassoula 2005b;
Martinez-Valpuesta \tal 2006). There are indications for it 
in the buckling strength, but not in the bar strength. As already
mentioned, there are several possible definitions of the bar and the
buckling strengths (Athanassoula \& Martinez-Valpuesta, in
preparation; Martinez-Valpuesta \& Athanassoula, these
proceedings) and exactly how clearly the buckling episodes are seen in
time evolution plots depends somewhat on the definition adopted. For
example, since the first buckling concerns more inner than 
outer parts (Martinez-Valpuesta \tal 2006), one can calculate the quantities in
Fig.~\ref{fig:tevl} using, instead of the whole disc, only the radial
range mainly concerned by the first or by the second buckling, which
allows to see better the buckling in question and its effects. 
   
\section{Velocity anisotropy and peanut formation}

As already mentioned, peanut formation is linked to the vertical
instability of parts of the main family of periodic orbits
constituting the bar. This 
family is widely known as the x$_1$ family. Its stability can be
followed from the stability diagram (see e.g. figures 3 and 4 of
Skokos \tal 2002)  
which show that, at the positions where x$_1$ becomes unstable, other
families bifurcate. These are linked to the $n:1$ vertical resonances
and extend well outside the disc equatorial plane. As shown by Patsis
\tal (2002), 
some of them are very good building blocks for the formation of
peanuts, because they are stable and because their orbits have the
right shape and extent. Thus, orbital structure theory can go a long
way towards explaining the formation and properties of the peanut.

\begin{figure}[!h]
\vspace{0.5cm}
\plotfiddle{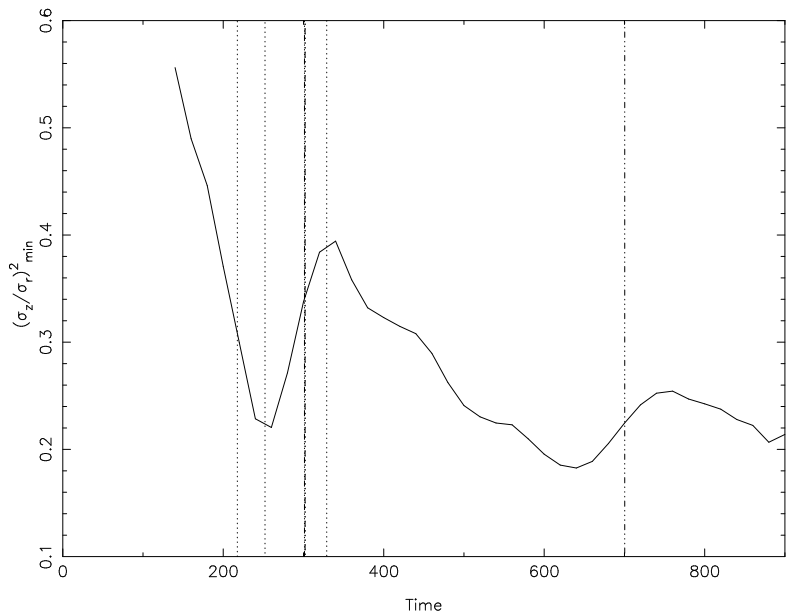}{7.3cm}{-90}{45}{45}{-190}{265}
\caption{Time evolution of the ratio $\sigma_z^2/\sigma_r^2$. The thin
  vertical lines mark characteristic times linked to bar
  formation and evolution. From left to right, the first four ones are
  the bar
  formation time, the maximum amplitude time, the bar decay time and
  the bar minimum amplitude time, all corresponding to the first
  buckling episode (see text and
  Fig.~\ref{fig:tevl}). The fifth and last vertical line
  (dash-dot-dot-dot) marks the time of the second buckling. 
} 
\label{fig:sigrat} 
\end{figure}

An alternative approach explains the buckling and peanut formation as
due to the bending, or fire-hose, instability, studied analytically in
the linear regime (Toomre 1966; Araki 1985). These studies assign 
a critical value of $R_{\sigma}=\sigma_z^2/\sigma_r^2$ igniting
the onset of the instability, which is around 0.1. A number of
simulations, however, have shown that this instability sets in already
at much larger values (e.g. Merritt \& Sellwood 1994; Sotnikova \&
Rodionov 2003).  

To test this hypothesis, we calculate the radial and $z$
components of the disc velocity dispersion  
as a function of radius (averaging over azimuth and height). We then
find the minimum value of their ratio $R_{\sigma}$ and
plot its time evolution in Fig.~\ref{fig:sigrat}. 
The thin vertical lines in this figure are at the locations
found from Fig.~\ref{fig:tevl} and mark the characteristic times linked
to bar formation and evolution; namely the bar
formation time, the maximum amplitude time, the bar decay time and
the bar minimum amplitude time. Their location is clearly linked to the
various evolutionary phases of $R_{\sigma}$. This, however, need not
necessarily be seen as the cause of the buckling, but can also be seen
as its result. Indeed, as the bar forms $\sigma_r$ increases
drastically, so that $R_{\sigma}$ decreases. Then, the bar
amplitude reaches its maximum and starts decreasing, while the peanut
starts forming. During this time, $\sigma_r$ decreases, while
$\sigma_z$ increases. As a result, between these two time intervals the
ratio $R_{\sigma}$ should reach a minimum and then increase again, as
is indeed seen in Fig.~\ref{fig:sigrat}. Then the
bar amplitude reaches a minimum, which corresponds to a minimum of
$\sigma_r$ and therefore to a maximum of $R_{\sigma}$. This is
followed by a slow decrease of $R_{\sigma}$, which is stopped by  
the second buckling 
episode. The value of $R_{\sigma}$ at which this instability sets in
is much less extreme than that predicted by the above mentioned
analytical works, but in good agreement with other $N$-body
simulations. 

More work is necessary before we fully understand the respective roles
of the orbital structure results and of the velocity anisotropy
effects on the formation and evolution of B/P structures. Both 
explain part of the story, but many aspects of their interplay are
still unclear. Orbital structure results can tell us whether the 
appropriate building blocks are available, or not, and this is
essential since the lack of the appropriate building blocks can
prohibit the formation of a given structure. Furthermore, studies of
the properties of the building-block orbits are essential for
understanding the properties of the B/P structures. On the
other hand, it is necessary to group all these building blocks into one
coherent unit and here collective effects are essential. They also can
place limits on the formation of B/P structures, as well as give
information on their properties. We will discuss the respective input
from the two methods further elsewhere.
 
\section{Bar and peanut destruction}

The effect of a central mass concentration (CMC) on bar evolution 
has been well studied with the help of purely stellar $N$-body simulations, 
placing lower limits on the mass and central concentration
necessary for the bar to be destroyed. Comparison with observations,
however, shows that these limits are much higher than the measured
values (Shen \& Sellwood 2004; Athanassoula, Lambert \& Dehnen 2005
and references therein), thus 
casting doubts as to whether it is possible to destroy bars in this
way. On the other hand, simulations with an SPH, or sticky particle
representation of the gas come to a different conclusion
(e.g. Friedli \& Benz 1993; Berentzen \tal 1998; Bournaud \& Combes
2002). Whether the difference 
between the two types of simulations could be due to
the role of the gas in the angular momentum exchange
is still debated (Bournaud, Combes \& Semelin 2005; 
Athanassoula \tal 2005; Berentzen \tal 2007). 
We will not attempt to settle the issue
here. We will just study the photometric properties
after the bar is destroyed, without worrying about whether
the necessary CMC is compatible with observations, or not, or what the
role of the gaseous component is. Our aim here is to see whether these
photometric properties are compatible with those of a bulge, and, if
yes, of what type of bulge. 

\begin{figure}[!ht]
\plotone{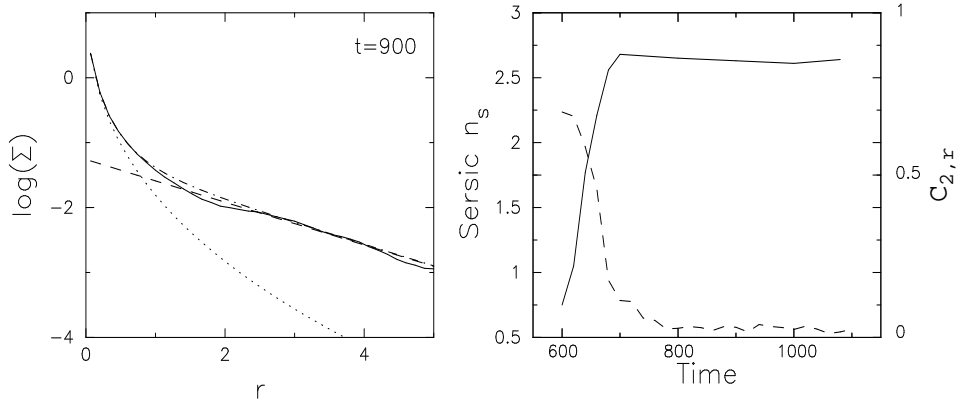}
\caption{Bulge formation via bar destruction (see text). {\it Left
    panel :} Radial density profile of the luminous material (solid
    line) and its decomposition in a disc (dashed) and bulge (dotted)
    component. The sum of the two is given by a dash-dotted line. {\it
    Right panel :} Time evolution of the S\'ersic index (solid line) and
    of the bar strength (dashed line). All quantities are in computer
    units as given by Athanassoula \tal (2005).  
} 
\label{fig:CMC} 
\end{figure}

The simulations used in this analysis are those of Athanassoula \tal (2005) 
For several times in every simulation, we obtained the radial density
profile,  and then decomposed it
into a disc and a bulge component, fitted
by an exponential disc and a S\'ersic profile, respectively. An example
for a case with initially a strong bar with a clear peanut, in which
we gradually grew a massive CMC, is shown in the left panel of 
Fig.~\ref{fig:CMC}. The corresponding
value for the S\'ersic index is $n_s$ = 2.6, i.e. approaching the
range of values found for classical bulges (e.g. Kormendy \& Kennicutt
2004). We have performed this 
decomposition for several times during the simulation and plot the
values of $n_s$ as a function of time in the right panel
of Fig.~\ref{fig:CMC}, where we also plot the bar strength as a function
of time. For the simulation we are analysing, the CMC has been grown
gradually in 100 computer units (600 - 700), or, 
equivalently, 1.4 Gyrs. We note that, during that time, the bar
strength drops and  $n_s$ increases, both
drastically, while after the CMC has reached its final mass, both
values stay nearly constant. The final value of the S\'ersic
index depends on the mass of the CMC.

The above shows that, at least as far as the photometrical definition
presented in Sect. 1 is concerned, a decayed bar can give rise to an
object with a S\'ersic profile approaching those of classical bulges.     
Its shape when the snapshot is viewed edge-on is compatible with that
of a classical bulge. Thus, by two of the three definitions
presented in Sect. 1, it is tempting to classify this object as a classical
bulge. Work to study its kinematics is underway, in order to complete
the classification and see whether such objects are compatible with
classical bulges (Athanassoula, Aguerri \&
Martinez-Valpuesta, in preparation).
    
\acknowledgements 
It is a pleasure to thank our collaborator, Alfonso Aguerri, 
for many interesting discussions and a fruitful collaboration, 
and A. Bosma, M. Balcells, P. Erwin,
C. Heller, R. Peletier, M. Pohlen and I. Shlosman
for stimulating discussions and
encouragement. This
work has been partially supported by a Peter Gr\"uber 
Foundation Fellowship to IMV and by grant ANR-06-BLAN-0172.


\begin{thebibliography}{}
\bibitem []{}Aguerri, J.~A.~L., Balcells,~M., Peletier, R.~F. 2001, A\&A, 367, 428
\bibitem[\protect\citeauthoryear{Araki}{1985}]{Araki}
Araki, S. 1985 PhD thesis, MIT
\bibitem []{}Athanassoula, E. 1992, MNRAS, 259, 345
\bibitem[\protect\citeauthoryear{Athanassoula}{1999}]{A99}Athanassoula,
  E. 1999, in {\it Astrophysical Discs}, eds. J. A.  
  Sellwood \& J. Goodman,  PASP conference series, 160, 351
\bibitem[\protect\citeauthoryear{Athanassoula}{2003}]{a03}
Athanassoula, E. 2003, MNRAS, 341, 1179 
\bibitem[\protect\citeauthoryear{Athanassoula}{2005a}]
  {a05a}
Athanassoula, E. 2005a, MNRAS, 358, 1477 
\bibitem[\protect\citeauthoryear{Athanassoula}{2005b}]
  {a05b} Athanassoula, E. 2005b, in 
{\it Planetary Nebulae as Astronomical Tools},
eds. R. Szczerba, G. Stasi\'nska \& S. K. G\'orny, AIP Conf. Proc.
804, Melville, New  York, 333
\bibitem[\protect\citeauthoryear{Athanassoula \& Misiriotis}{2002}]{am02}
Athanassoula, E., Misiriotis, A. 2002, MNRAS, 330, 35 
\bibitem []{}Athanassoula, E., Lambert, J.~C., Dehnen, W. 2005, MNRAS, 363, 496 
\bibitem []{Ba+2001} Bacon, R. \tal 2001, MNRAS, 326, 23
\bibitem []{}Berentzen, I., Heller, C. H., Shlosman, I., Fricke, K. J. 1998,
MNRAS, 300, 49
\bibitem []{}Berentzen, I., Shlosman, I., Martinez-Valpuesta, I., Heller, C.H.
     2007, ApJ, in press 
\bibitem []{}Binney, J. 1978, MNRAS, 183, 501
\bibitem []{}Binney, J. 2005, MNRAS, 363, 937
\bibitem []{}Bournaud, F., Combes, F. 2002, A\&A, 392, 83
\bibitem []{}Bournaud, F., Combes, F., Semelin, B. 2005, MNRAS, 364L, 18
\bibitem[\protect\citeauthoryear{Combes \& Sanders}{1981}]{CombesSanders81}
Combes, F., Sanders, R. H. 1981, A\&A, 96, 164
\bibitem[\protect\citeauthoryear{Combes \tal}{1990}]{CDFP90}
Combes, F., Debbasch, F., Friedli, D., Pfenniger, D. 1990, A\&A,
  233, 82 
\bibitem []{}Debattista, V.~P., Carollo, M., Mayer, L., Moore, B. 2004, ApJ, 604, L93
\bibitem []{}Debattista, V.~P., Carollo, M., Mayer, L. Moore, B., Wadsley, J., Quinn, T. 2006, ApJ, 645, 209
\bibitem []{} de Zeeuw, P.~T. \tal 2002, MNRAS, 329, 513
\bibitem []{}Friedli, D., Benz, W. 1993, A\&A, 268, 65
\bibitem []{}Fu, Y.~N., Huang, J.~H., Deng, Z.~G. 2003, MNRAS, 339, 442
\bibitem []{}Heller, C.~H., Shlosman, I. 1994, ApJ, 424, 84
\bibitem []{}Immeli, A., Samland, M., Gerhard, O., Westera, P. 2004, A\&A, 413, 547
\bibitem []{}Kormendy, J. 1993, in {\it Galactic Bulges}, eds. H. Dejonghe and
  H. J. Habing, Kluwer Academic Publ., IAU Symposium 153, 209
\bibitem []{}Kormendy J., Kennicutt R.~C. 2004, ARA\&A, 42, 603
\bibitem [\protect\citeauthoryear{Martinez-Valpuesta \& Shlosman}
{2004}]{MartShlo1} 
Martinez-Valpuesta, I., Shlosman, I. 2004, ApJ, 613, 29
\bibitem [\protect\citeauthoryear{Martinez-Valpuesta, Shlosman \& Heller}
{Martinez-Valpuesta \tal}{2006}]{MartShlo2} Martinez-Valpuesta, I., Shlosman, I., Heller, C. 2006, ApJ, 637, 214
\bibitem [{Merritt \& Sellwood}{1994}]{MeSel94} Merritt, D., Sellwood J.~A. 1994, ApJ, 425, 551
\bibitem [{Noguchi}{1998}]{Noguchi}Noguchi, M. 1998, Nature, 392, 253
\bibitem[{O'Neil /& Dubinski}{2003}]{OnDu03} O'Neill, J.~K. Dubinski, J. 2003, MNRAS,346, 251
\bibitem[\protect\citeauthoryear{Patsis Skokos \& Athanassoula}{Patsis \tal}{2002}]{PSA02}
Patsis, P., Skokos, Ch., Athanassoula, E. 2002, MNRAS, 337, 578 
\bibitem []{}Pfenniger, D. 1984, A\&A, 134, 373
\bibitem []{}Pfenniger, D. 1991, in {\it Dynamics of Disc Galaxies},
  ed. B. Sundelius, G\"oteborg University Press, G\"oteborg 
\bibitem[\protect\citeauthoryear{Raha \tal}{1991}]{RSJK91}
Raha, N., Sellwood, J. A., James, R. A., Kahn, F. D. 1991,
  Nature, 352, 411  
\bibitem[\protect\citeauthoryear{S\'ersic}{1968}]{Sersic}
S\'ersic, J. 1968, Atlas de Galaxias Australes, Obs. Astron. Cordoba
\bibitem []{} Shen, J., Sellwood, J.~A. 2004 ApJ, 604, 614
\bibitem[\protect\citeauthoryear{Skokos Patsis \& Athanassoula}{Skokos \tal}{2002a}]{SPAa}
Skokos, H., Patsis, P., Athanassoula, E. 2002, MNRAS, 333, 847 
\bibitem []{}Sotnikova, N.~Ya., Rodionov, S.~A. 2003, AstL, 29, 321
\bibitem []{}Steinmetz, M., M\"uller, E. 1995, MNRAS, 276, 549
\bibitem[\protect\citeauthoryear{Toomre}{1966}]{Toomre}
Toomre, A. 1966, Geophys. Fluid Dyn., N66-46, 111
\bibitem [{Wada \& Habe}{1995}]{WaHa95}Wada, K., Habe, A. 1995, MNRAS, 277, 433
\end{thebibliography}
\end{document}